\documentclass[a4paper,12pt]{article}
\usepackage{amssymb,amsmath,bm}
\usepackage{mathrsfs}
\usepackage{graphicx}
\usepackage{enumerate}

\title{Utilitarian Theorems and Equivalence of Utility Theories}
\author{Yuhki Hosoya\thanks{TEL: +81-90-5525-5142, E-mail: ukki(at)gs.econ.keio.ac.jp}\\ Faculty of Economics, Chuo University\thanks{742-1, Higashinakano, Hachioji-shi, Tokyo, 192-0393, Japan.}}
\date{}
\begin{document}
\maketitle

\begin{abstract}
In this paper, we consider an environment in which the utilitarian theorem for the NM utility function derived by Harsanyi and the utilitarian theorem for Alt's utility function derived by Harvey hold simultaneously, and prove that the NM utility function coincides with Alt's utility function under this setup. This result is so paradoxical that we must presume that at least one of the utilitarian theorems contains a strong assumption. We examine the assumptions one by one and conclude that one of Harsanyi's axioms is strong.

\vspace{12pt}

\vspace{12pt}
\end{abstract}

\section{Introduction}
Utilitarianism is the idea that each individual's evaluation of the state of a society can be represented by some quantity that is called utility, and that the higher the sum of utilities, the more desirable the state of the society. This idea was strongly promoted by Jeremy Bentham in the 19th century, but has long been criticized for the ambiguity of the criteria for how the actual utility value is calculated. According to the utilitarian view, the amount of change in utility associated with a change in the state of the society must express the intensity of the improvement for this person, but there was no theory in economics in the 19th century that could deal with a function expressing the strength of improvement. This problem was mathematically formulated by Alt (1936), who derived an existence theorem for the utility function that represents the intensity of improvement using an axiomatic approach. This work was refined by Krantz et al. (1971) and later developed by Shapley (1975) and Hosoya (2022).

On the other hand, Harsanyi (1955) derived a theorem mathematically endorsing the measurement of the goodness of the state of a society by the sum of its utility in a completely different context. He first argued that, because society is always faced with uncertainties, the model must include stochastic variations in order to evaluate the goodness of the state of the society. He then assumed that individual and social preferences are defined on the space of simple lotteries, and derived the result that if only three axioms seem to be weak hold, the preference of the society must be expressed as a weighted sum of individual NM utility functions. This result is known as Harsanyi's utilitarian theorem.

Because the message of Harsanyi's theorem was so radical, his result provoked a variety of reactions, including some emotional ones. First, although his result was purely mathematical, he also discussed at length in the same paper the problems in conducting a utilitarian analysis, and even discussed the NM utility function there as if it were an Alt's utility function. With Harsanyi's argument in mind, Luce and Raiffa (1957) argued that the NM utility function should not be given an interpretation as an Alt's utility function. In contrast, according to researchers who favor Harsanyi's result, such criticism is invalid because Harsanyi's result is purely mathematical, and does not represent an ideology. Sen opposed Harsanyi particularly strongly in this context. He apparently considered the exclusion of Rawls' maximin principle from Harsanyi's formulation to be a very serious problem. As a result, Harsanyi and Sen engaged in several debates in the journal ``Theory and Decision''. Representative references include Harsanyi (1975) and Sen (1976).

On the other hand, Harvey (1999) derived another utilitarian result for Alt's utility function. Specifically, he showed that, under certain axioms, the social preference is expressed in terms of a weighted sum of individual Alt's utility functions. Similar to Harsanyi's result, Harvey's axiom does not appear to be strong at first glance.

In this paper, we first analyze what happens in an environment where both Harsanyi's and Harvey's utilitarian theorems simultaneously hold in order to test whether Luce--Raiffa's criticism is valid, and induce a theorem stating that, in this circumstance, Alt's utility and NM utility must coincide for all individuals. Thus, in this sense, the discourse ``Harsanyi's result is mathematical, and thus Luce--Raiffa's criticism is off the mark'' has some validity.

From an optimistic view, this result is excellent. If all our axioms are accepted, it is no longer necessary to distinguish between NM utility functions and Alt's utility functions, and thus, for example, any method for estimating an NM utility function can be used directly for Alt's utility function. This can be seen as a result of eliminating interpretive difficulty for many applied studies. 

However, this result can also be viewed as a pesimistic result. To the best of our understanding, Alt's utility theory and NM utility theory appear to be independent and unrelated. Moreover, at first glance, neither Harsanyi's nor Harvey's theorem seems to have any axiom that would bridge the difference between the two. The fact that the two different utility concepts nevertheless coincide suggests that there is perhaps an axiom within two utilitarian theorems that appears weak at first glance, but is in fact very strong. We continue our analysis, and finally find that one axiom in Harsanyi's utilitarian theorem is much stronger than would normally be thought.

At least one of the axioms in Harsanyi's utilitarian theorem is so strong that it makes two very different notions of utility functions coincide. On the other hand, the concepts of Alt's utility and NM utility are so different. If these two utility functions should not coincide, it follows that Harsanyi's axiom may be inadequate. This argument can be appreciated as a modern refinement of Luce--Raiffa's critique.

Our main result is a congruence theorem between Alt's utility and NM utility, which is presented as Theorem 3 in this paper. In addition to this result, this paper contains variants of both utilitarian theorems (Theorems 1-2).

This paper is organized as follows. Section 2 contains several preliminaries for our main result. In this section, we explain the basic results on NM utility and Alt's utility, and introduce the mathematical formulation of ``society''. We also explain two utilitarian theorems. Section 3 addresses the main result and its proof. Following this main result, Section 4 provides an in-depth analysis of why the result holds. Section 5 contains the conclusion. The proofs of all results are contained in the appendix.

\section{Preliminaries}
\subsection{Weak Order and NM Utility Functions}
In this subsection, we present basic knowledge on weak orders. First, consider a nonempty set $X$. A subset $\succsim$ of $X^2$ is called a {\bf binary relation} on $X$. A binary relation $\succsim$ on $X$ is called a {\bf weak order} on $X$ if and only if, 1) (completeness) either $(x,y)\in \succsim$ or $(y,x)\in \succsim$ for every $x,y\in X$, and 2) (transitivity) if $(x,y)\in \succsim$ and $(y,z)\in \succsim$, then $(x,z)\in \succsim$. If $\succsim$ is a weak order on $X$, then we write $x\succsim y$ instead of $(x,y)\in \succsim$. Moreover, we write $x\succ y$ if and only if $x\succsim y$ and $y\not\succsim x$, and $x\sim y$ if and only if $x\succsim y$ and $y\succsim x$.

Suppose that $\succsim$ is a weak order on $X$. If a function $u:X\to \mathbb{R}$ satisfies the following relationship:
\[x\succsim y\Leftrightarrow u(x)\ge u(y),\]
then $u$ is said to {\bf represent} $\succsim$.

Suppose that $\mathscr{P}$ is the set of all probability measures with a finite support on a nonempty set $X$. We call an element of $\mathscr{P}$ a {\bf simple lottery} on $X$. For any function $f:X\to \mathbb{R}$ and $P\in \mathscr{P}$, define $E_P[f]$ as the expectation of $f$ with respect to $P$: that is, if the support of $P$ is $\{x_1,...,x_m\}$, then
\[E_P[f]=\sum_{i=1}^mP(\{x_i\})f(x_i).\]
Suppose that $\succsim$ is a weak order on $\mathscr{P}$. Then, it is said to be {\bf independent} if and only if for every $P,Q\in\mathscr{P}$, $P\succ Q$ implies that
\[(1-t)P+tR\succ (1-t)Q+tR\]
for every $R\in \mathscr{P}$ and every $t\in [0,1[$. Moreover, it is said to be {\bf segmentally continuous} if and only if for every $P,Q,R\in \mathscr{P}$ such that $P\succ Q\succ R$, the following two sets
\[\{t\in [0,1]|(1-t)P+tR\succsim Q\},\ \{t\in [0,1]|Q\succsim (1-t)P+tR\}\]
are closed.

Let $\delta_x$ denote the Dirac measure whose support is $\{x\}$. Choose any weak order $\succsim$ on $X$. We say that a weak order $\succeq$ on $\mathscr{P}$ is a {\bf probabilistic extension} of $\succsim$ if and only if the following relationship holds:
\[x\succsim y\Leftrightarrow \delta_x\succeq \delta_y.\]
The following theorem is well known.\footnote{This theorem was originally proved by von Neumann and Morgenstern (1944). For a modern proof, see ch.5 of Kreps (1988).}

\vspace{12pt}
\noindent
{\bf Theorem}. Suppose that $\mathscr{P}$ is as defined above and $\succsim$ is a weak order on $\mathscr{P}$. Then, the following two statements are equivalent.
\begin{enumerate}[1)]
\item $\succsim$ is independent and segmentally continuous.

\item There exists a function $u:X\to \mathbb{R}$ such that for $U(P)=E_P[u]$, $U$ represents $\succsim$.
\end{enumerate}
Moreover, such a $u$ is unique up to a positive affine transformation.

\vspace{12pt}
We call the above $u$ a {\bf Neumann--Morgenstern expected utility function}, or abbreviately, an {\bf NM utility function} of $\succsim$.\footnote{Actually, $U(P)=E_P[u]$ is the main body of the utility function, and thus in some textbooks, $U(P)$ is called the NM utility function, and they call $u$ some different name, for example, utility index or utility kernel. In this paper, however, we want to compare this function with Alt's utility function, whose domain is not $\mathscr{P}$ but $X$. Therefore, we call $u$ the `NM utility function' in this paper. We believe that this does not make any confusion.}

\subsection{Alt's System}
Let $X$ be a nonempty set, and suppose that $\ge$ is a weak order on $X^2$. We call this $\ge$ an {\bf Alt's system} on $X$. We write $[x,y]\ge [z,w]$ instead of $(x,y,z,w)\in \ge$. Moreover, we write $[x,y]>[z,w]$ if $[x,y]\ge [z,w]$ and $[z,w]\not\ge [x,y]$, and $[x,y]=[z,w]$ if $[x,y]\ge [z,w]$ and $[z,w]\ge [x,y]$.

Note that, an Alt's system $\ge$ represents the intensity of improvement. That is, $[x,y]\ge [z,w]$ implies that the intensity of improvement associated with moving from $y$ to $x$ is stronger than or equal to that associated with moving from $w$ to $z$.

A function $u:X\to \mathbb{R}$ is said to {\bf represent} $\ge$ if and only if
\[[x,y]\ge [z,w]\Leftrightarrow u(x)-u(y)\ge u(z)-u(w)\]
for every $x,y,z,w\in X$, and in this case, we call $u$ an {\bf Alt's utility function} of $\ge$.

Let $\succsim$ be a weak order on $X$ and $\ge$ be an Alt's system on $X$. We say that $\succsim$ {\bf matches} $\ge$ (or, $\ge$ matches $\succsim$) if and only if for every $x,y\in X$,
\[x\succsim y\Leftrightarrow [x,y]\ge [y,y].\]
Suppose that $u$ represents $\ge$ and $\succsim$ matches $\ge$. Then,
\[x\succsim y\Leftrightarrow [x,y]\ge [y,y]\Leftrightarrow u(x)-u(y)\ge u(y)-u(y)\Leftrightarrow u(x)\ge u(y),\]
and thus Alt's utility function $u$ also represents $\succsim$.

We introduce three axioms on Alt's system. An Alt's system $\ge$ satisfies {\bf consistency} if and only if for every $x,y,z\in X$,
\[[x,y]\ge [y,y]\Leftrightarrow [x,z]\ge [y,z].\]
Next, an Alt's system $\ge$ satisfies the {\bf crossover axiom} if and only if for every $x,y,z,w\in X$,
\[[x,y]=[z,w]\Leftrightarrow [x,z]=[y,w].\]
Third, suppose that $X$ is a topological space. An Alt's system $\ge$ satisfies {\bf continuity} if and only if it is closed in $X^4$.

Hosoya (2022) showed the following result.

\vspace{12pt}
\noindent
{\bf Theorem}. Suppose that $X$ is a separable and path-connected Hausdorff topological space, and that $\ge$ is an Alt's system on $X$. Then, there exists a continuous representation $u:X\to \mathbb{R}$ of $\ge$ if and only if $\ge$ satisfies consistency, the crossover axiom, and continuity. Moreover, in this case, $u$ is unique up to a positive affine transformation.

\subsection{Societies}
Fix any $n\ge 2$, and let $X$ be a nonempty set. We call an $(n+1)$-tuple $(\succsim_0,...,\succsim_n)$ of weak orders on $X$ a {\bf society} on the state space $X$. The weak order $\succsim_0$ is treated as representing an {\bf ethical} order for this society, and for $i\in \{1,...,n\}$, $\succsim_i$ represents the $i$-th person's individual preference.

Let a society $(\succsim_0,...,\succsim_n)$ on $X$ be given. Choose any $x,y\in X$. We say that $y$ is Pareto-dominated by $x$ if, 1) $x\succsim_iy$ for each $i\in \{1,...,n\}$ and 2) $x\succ_iy$ for some $i\in \{1,...,n\}$. We say that the society $(\succsim_0,...,\succsim_n)$ obeys the {\bf Pareto criterion} if for every pair $(x,y)$ such that $y$ is Pareto-dominated to $x$, $x\succ_0y$.

Let $\mathscr{P}$ be the set of all simple lotteries on $X$. Choose any society $(\succsim_0,...,\succsim_n)$ on $X$. We call a society $(\succeq_0,...,\succeq_n)$ on $\mathscr{P}$ a {\bf probabilistic extension} of $(\succsim_0,...,\succsim_n)$ if and only if for every $i\in \{0,...,n\}$, $\succeq_i$ is a probabilistic extension of $\succsim_i$.

The following theorem is an extension of the main theorem of Harsanyi (1955).\footnote{Weymark (1993) showed this result under the finiteness assumption of $X$.}

\vspace{12pt}
\noindent
{\bf Theorem 1}. Let $X$ be a nonempty set and $\mathscr{P}$ be the set of all simple lotteries on $X$. Choose a society $(\succsim_0,...,\succsim_n)$ on $\mathscr{P}$. Suppose that there exists an NM utility function $v:X\to \mathbb{R}$ of $\succsim_0$, and for any $i\in \{1,...,n\}$, there exists an NM utility function $u_i:X\to \mathbb{R}$. Then, the following two properties are equivalent.
\begin{enumerate}[(i)]
\item If $P\sim_iQ$ for all $i\in \{1,...,n\}$, then $P\sim_0Q$.

\item There exist $a_1,...,a_n\in \mathbb{R}$ and $b\in \mathbb{R}$ such that
\begin{equation}
v(x)=\sum_{i=1}^na_iu_i(x)+b
\end{equation}
for all $x\in X$.
\end{enumerate}
Moreover, this society obeys the Pareto criterion if and only if we can choose $a_1,...,a_n,b$ so that $a_i>0$ for all $i\in \{1,...,n\}$.\footnote{Note that, if $(u_i)$ is not independent, then $a_1,...,a_n$ is not determined uniquely, and thus ``$a_i>0$ must hold'' is wrong even when the society obeys the Pareto criterion.}

\vspace{12pt}
We say that a society on $X$ is {\bf separable} if, 1) $X=\prod_{i=1}^nX_i$, and 2) for any $x=(x_1,...,x_n)$ and $y=(y_1,...,y_n)$, if $x_i=y_i$, then $x\sim_iy$. On the other hand, we say that a society is {\bf semi-separable} if, for any $x_1,...,x_n\in X$, there exists $x\in X$ such that $x_i\sim_ix$ for any $i\in \{1,...,n\}$. This assumption is needed for later arguments. It is easy to show that any separable society is semi-separable.

\vspace{12pt}
\noindent
{\bf NOTE}: At first glance, the assumption of semi-separability may seem artificial and undesirable. However, quite a few models commonly used in economics satisfy this assumption. Let us look at some examples.

The first example is a model of a pure exchange economy in the general equilibrium theory. Each individual $i$ has a consumption set $\Omega_i$ and an order $\succsim_i$ on this set. The state of the society is represented by a consumption profile $(x_1,...,x_n)$. Thus, $X=\prod_{i=1}^n\Omega_i$, and if $i\neq j$, then $\succsim_i$ is not affected by $x_j$. This society is separable.

As a second example, we consider a general equilibrium model with a production structure. Each firm $j$ has a production set $Y_j$. Therefore, the space of the state of the society $X$ is represented by the following set:
\[\prod_{i=1}^n\Omega_i\times \prod_{j=1}^mY_j.\]
Because $Y_j$ is included, this society is not separable. However, it is semi-separable. For example, if $z^i=(x_1^i,...,x_n^i,y_1^i,...,y_m^i)\in X$, then for $z=(x_1^1,...,x_n^n,y_1^1,...,y_m^1)\in X$, $z^i\sim_iz$ for each $i$.

As this example shows, the problem of whether a society is semi-separable is not affected by the feasibility problem. In the example above, even if each $z^i$ is feasible, $z$ is not necessarily feasible. However, this has nothing to do with the problem of whether this society is semi-separable.

Thus, societies including serious externalities can also be semi-separable. For example, let us consider a pure exchange economy with externalities. In this economy, the order $\succsim_i$ of individual $i$ is defined on $\prod_{j=1}^n\Omega_j$ insread of $\Omega_i$. Let us introduce side-payment into the model, and consider $t_i\in \mathbb{R}$ as a government subsidy if it is positive and a tax if it is negative. The state of the society is represented by $(x_1,...,x_n,t_1,...,t_n)$, and if $u_i$ represents $\succsim_i$, assume that $U_i(x,t_i)=u_i(x)+t_i$ represents the preference of $i$. Then, it is easy to check that this society is semi-separable. Again, there is no need to discuss the financial resources of $t_i$. Even if there is a profile $(t_1,...,t_n)$ that is not feasible, it would have no effect on whether the society is semi-separable.

As well as externalities, a society with public goods can be semi-separable insofar as side-payments are allowed. Thus, many models satisfy the condition of semi-separability, and this assumption does not seem to be strong.

\subsection{Social Systems}
Let $X$ be a nonempty set. We call an $(n+1)$-tuple $(\ge_0,...,\ge_n)$ of Alt's systems on $X$ a {\bf social system} on the state space $X$. As in the case of societies, the Alt's system $\ge_0$ is treated as representing an ethical intensity of improvement, and for $i\in \{1,...,n\}$, $\ge_i$ represents the $i$-th person's individual intensity of improvement.

Choose a society $(\succsim_0,...,\succsim_n)$ and a social system $(\ge_0,...,\ge_n)$ on $X$. We say that this society {\bf matches} this social system (or, this social system matches this society) if and only if $\succsim_i$ matches $\ge_i$ for all $i\in \{0,...,n\}$.

The following theorem is a variant of the main result of Harvey (1999).

\vspace{12pt}
\noindent
{\bf Theorem 2}. Suppose that $X$ is a connected Hausdorff topological space, and a semi-separable society $(\succsim_0,...,\succsim_n)$ on $X$ matches a social system $(\ge_0,...,\ge_n)$ on $X$. Moreover, suppose that this society obeys the Pareto criterion, that a continuous function $v:X\to \mathbb{R}$ represents $\ge_0$, and that a continuous function $u_i:X\to \mathbb{R}$ represents $\ge_i$ for each $i\in \{1,...,n\}$. Then, the following two properties are equivalent.
\begin{enumerate}[(I)]
\item If $[x,y]=_i[z,w]$ for all $i\in \{1,...,n\}$, then $[x,y]=_0[z,w]$.

\item There exist $a_1,...,a_n\in \mathbb{R}_{++}$ and $b\in \mathbb{R}$ such that
\begin{equation}
v(x)=\sum_{i=1}^na_iu_i(x)+b
\end{equation}
for all $x\in X$.
\end{enumerate}

\subsection{Notes on the Continuity of Utility Functions}
Later, we require the continuity of the NM utility function and Alt's utility function. Hence, in this subsection, we discuss this topic.

In this subsection, we consider that $X$ is a topological space, and $\mathscr{P}$ is the set of all simple lotteries on $X$. Let $\succsim$ be a weak order on $\mathscr{P}$. If $\succsim$ is independent and segmentally continuous, then there exists an NM utility function $u:X\to \mathbb{R}$ of $\succsim$. We want to obtain a necessary and sufficient condition for $u$ to be continuous.

In this context, continuity is sometimes mentioned as a necessary and sufficient condition for the NM utility function to be {\bf bounded} and continuous. See, for example, Foldes (1972), Grandmont (1972), and ch.5 of Kreps (1988). However, in this context, $X$ is usually treated as a metric space, and $\mathscr{P}$ is usually the space of all probability measures on $X$. In our model, $X$ is not necessarily a metric space, and $\mathscr{P}$ is the set of simple lotteries. This implies that their results cannot be applied directly.

Nevertheless, we expect that a similar result holds for our setups. Indeed, we can prove that $u$ is continuous if $\succsim$ is closed with respect to the weak* topology (explained later) on $\mathscr{P}^2$. However, in this case, $u$ must be bounded. The boundedness of the NM utility function is undesirable for our later arguments. Although several studies such as Dillenberger and Krishna (2014) replace the boundedness with some kind of growth condition, these results are also undesirable for us. Therefore, we make a slightly different argument. 

First, we prove the following lemma.

\vspace{12pt}
\noindent
{\bf Lemma 1}. Suppose that $X$ is a Tychonoff space,\footnote{A Hausdorff topological space $X$ is called a {\bf Tychonoff space} if and only if for any $x\in X$ and a closed set $A\subset X$ such that $x\notin A$, there exists a continuous function $f:X\to [0,1]$ such that $f(x)=1$ and $f(y)=0$ for all $y\in A$. Note that, if a Hausdorff topological space $X$ is normal, then by Urysohn's theorem, it is a Tychonoff space. In particular, any metric space is a Tychonoff space.} and that $\mathscr{P}$ is the set of all simple lotteries on $X$. Then, there exists a Hausdorff topology such that a net $(P_{\nu})$ on $\mathscr{P}$ converges to $P\in \mathscr{P}$ in this topology if and only if for every bounded and continuous function $u:X\to \mathbb{R}$, the net $(E_{P_{\nu}}[u])$ converges to $E_P[u]$.

\vspace{12pt}
We call this topology the {\bf weak* topology}.

Let $\succsim$ be a weak order on $\mathscr{P}$. Choose any $Y\subset X$, and define $\mathscr{P}_Y$ as the set of all element $P\in\mathscr{P}$ whose support is included in $Y$. Define $\succsim_Y=\succsim\cap (\mathscr{P}_Y)^2$. We call this $\succsim_Y$ the {\bf restriction} of $\succsim$ to $Y$. We say that $\succsim$ is {\bf sequentially continuous} if there exists a sequence $(X_n)$ of closed subsets of $X$ such that 1) for each $n$, $X_n$ is included in the interior of $X_{n+1}$, 2) $\cup_nX_n=X$, and 3) for each $n$, $\succsim_{X_n}$ is closed in $(\mathscr{P}_{X_n})^2$ with respect to the weak* topology. Then, the following result holds.

\vspace{12pt}
\noindent
{\bf Lemma 2}. Suppose that $X$ is a Tychonoff space, $\mathscr{P}$ is the set of all simple lotteries on $X$, and $\succsim$ is a sequentially continuous weak order on $\mathscr{P}$. Then, $\succsim$ is segmentally continuous. In particular, if $\succsim$ is a weak order on $\mathscr{P}$, then it is independent and sequentially continuous if and only if there exists a continuous NM utility function $u$ of $\succsim$.

\vspace{12pt}
Next, suppose that $\ge$ is an Alt's system on $X$ that has a continuous representation $u;X\to \mathbb{R}$. Hosoya (2022) showed that such a {\bf continuous} representation is unique up to positive affine transform. However, the ``continuous'' requirement can be dropped, and the following result holds.

\vspace{12pt}
\noindent
{\bf Lemma 3}. Suppose that $X$ is a connected Hausdorff topological space, $\ge$ is an Alt's system, and $u$ is a continuous representation of $\ge$. Then, any representation of $\ge$ must be a positive affine transform of $u$.

\section{Main Result}
Our main theorem is as follows.

\vspace{12pt}
\noindent
{\bf Theorem 3}. Let $X$ be a connected Hausdorff topological space. Choose any semi-separable society $(\succsim_0,...,\succsim_n)$ on $X$ that obeys the Pareto criterion. Suppose that this society has a probabilistic extension $(\succeq_0,...,\succeq_n)$ and a matching social system $(\ge_0,...,\ge_n)$. Suppose also that for each $i\in \{0,...,n\}$, $\succeq_i$ has a continuous NM utility function, and $\ge_i$ has a continuous Alt's utility function. Moreover, suppose that $(\succeq_0,...,\succeq_n)$ satisfies axiom (i) of Theorem 1, and $(\ge_0,...,\ge_n)$ satisfies axiom (I) of Theorem 2. Furthermore, suppose that there exist at least two people in $\{1,...,n\}$ such that $\succsim_i\neq X^2$. Then, for each $i\in \{1,...,n\}$, any NM utility function of $\succeq_i$ is an Alt's utility function of $\ge_i$, and vice versa.

\vspace{12pt}
We explain the idea of this theorem by introducing an example. Suppose that $n=2$, and $X=\mathbb{R}^2_+$. Moreover, suppose that the society is separable, where $X_i=\mathbb{R}_+$. For $i\in \{1,2\}$, let $u_i^*(x_i)$ (resp. $u_i(x_i)$) be the NM utility function (resp. Alt's utility function) of $i$. Moreover, let $v^*(x_1,x_2)$ (resp. $v(x_1,x_2)$) be the NM utility function (resp. Alt's utility function) of the ethical order. Suppose that both $u_i^*$ and $u_i$ represent $\succsim_i$, and both $v^*$ and $v$ represent $\succsim_0$. For simplicity, we assume that $v^*(x_1,x_2)=u_1(x_1)+u_2(x_2)$ and $v(x_1,x_2)=u_1(x_1)+u_2(x_2)$. Then, all requirements in Theorem 3 are satisfied, except the requirement that there exist at least two people such that $\succsim_i\neq X^2$.

Now, we assume that $u_1^*(x_1)=\sqrt{x_1}$ and $u_1(x_1)=x_1$. If $\succsim_2=X^2$, then there is no contradiction. Suppose that $\succsim_2\neq X^2$. Then, there exists $x_2^1,x_2^2\in \mathbb{R}_+$ such that $u_2^*(x_2^1)<u_2^*(x_2^2)$. Define $\varepsilon=u_2^*(x_2^2)-u_2^*(x_2^1)$, and let $x_1^k=(k\varepsilon)^2$. Then,
\begin{align*}
v^*(x_1^{k+1},x_2^1)=&~u_1^*(x_1^{k+1})+u_2^*(x_2^1)\\
=&~(k+1)\varepsilon+u_2^*(x_2^1)\\
=&~k\varepsilon+u_2^*(x_2^2)\\
=&~u_1^*(x_1^k)+u_2^*(x_2^2)=v^*(x_1^k,x_2^2),
\end{align*}
which implies that $(x_1^{k+1},x_2^1)\sim_0(x_1^k,x_2^2)$. Hence,
\[u_1(x_1^{k+1})+u_2(x_2^1)=v(x_1^{k+1},x_2^1)=v(x_1^k,x_2^2)=u_1(x_1^k)+u_2(x_2^2),\]
and thus,
\[(2k+1)\varepsilon^2=u_1(x_1^{k+1})-u_1(x_1^k)=u_2(x_2^2)-u_2(x_2^1),\]
which is impossible because the left-hand side depends on $k$ but the right-hand side is independent of $k$. Therefore, the possibility of $u_1^*(x_1)=\sqrt{x_1}$, $u_1(x_1)=x_1$ is rejected by the assumption $\succsim_2\neq X^2$. Actually, in this case, the only possibility is that $u_1^*(x_1)$ is a positive affine transform of $u_1(x_1)$, which is the claim of Theorem 3.

\vspace{12pt}
\noindent
{\bf NOTE}: As can be seen by analogy from the above explanation, this theorem does not use the properties of NM utility and Alt's utility at all. Therefore, the theorem can be more generalized. Suppose that there is a third utility concept different from both NM utility and Alt's utility, and that a utilitarian theorem on this concept holds. Then, Theorem 3 can be derived by replacing it with Alt's utility or NM utility. Hence, the third utility must also be identical to NM utility and Alt's utility. This point will be more clearly understood by looking at the proof in the appendix: see Proposition 1.

In this sense, the theorem need not be regarded as a theorem restricted to NM and Alt's utilities. However, the semi-separability is a peculiar assumption to Harvey's utilitarian theorem. And essentially, semi-separability is crucial to Theorem 3, and this theorem does not hold if this property is violated. To demonstrate this, we introduce an example where this theorem does not hold even if $u_1^*$ and $u_2^*$ are linearly independent.

Let $X=\{x\in \mathbb{R}^2_+|x_1+x_2=1\}$, and assume that $u_1(x)=\sqrt{x_1},\ u_2(x)=x_1, u_1^*(x)=x_1^2, u_2^*(x)=1-x_2^2$. Because $x_2=1-x_1$, $u_i$ represents the same preference as $u_i^*$. Moreover, both $(u_1,u_2)$ and $(u_1^*,u_2^*)$ are linearly independent. Let $v(x)=u_1(x)+u_2(x)$ and $v^*(x)=u_1^*(x)+u_2^*(x)$. Then,
\[v(x)=\sqrt{x_1}+x_1,\ v^*(x)=x_1^2-x_2^2+1=x_1-x_2+1=2x_1\]
and thus, $v$ represents the same preference as $v^*$. However, $u_i$ is not an affine transform of $u_i^*$, and thus the claim of Theorem 3 is violated in this case. It is easy to show that, in this case, among assumptions of Theorem 3, only semi-separability is violated.

\section{Discussion}
From Theorem 3, we conclude that two different notions of utility functions, the NM utility function and Alt's utility function, coincide in an environment where Harsanyi's and Harvey's utilitarian theorems hold. As we noted in the introduction, from an optimistic view, this result is good. Despite being interpretatively useful for analysis of economic welfare, Alt's utility theory has been criticized in that there is no method to estimate, or to reveal it. On the other hand, there are a number of studies on the estimation of the NM utility function. If the two utility concepts coincide, then in order to estimate Alt's utility, we only need to estimate NM utility. Thus, insofar as we accept all assumptions of Theorem 3, we can eliminate one major problem with Alt's utility. This is a good result for many applied studies.

However, from a pessimistic viewpoint, this result seems to be paradoxical. As already mentioned, the NM utility function is a concept that is related to probability, whereas Alt's utility function is a concept related to the intensity of improvement. These two concepts were borne from completely different ideas. Besides, neither axiom (i) of Theorem 1 nor axiom (I) of Theorem 2 appears to let the NM utility function express the intensity of improvement, nor does it allow Alt's utility function to reflect a probabilistic evaluation. If these coincide, we must suspect that some ridiculously strong requirement is mixed in axioms that do not appear strong on the surface.

Thus, let us analyze the setup of Theorem 3 in detail and consider which assumption can be regarded as strong.

\subsection{Existence and Properties of Probabilistic Extensions}
The first issue to consider is the existence of a probabilistic extension $\succeq_i$ of the preference relation $\succsim_i$, and the two axioms of independence and continuity imposed on $\succeq_i$. First, the independence axiom is often the subject of criticism when discussing expected utility. It is well-known that this axiom fails to explain the counterexample of Ellsberg (1961). To remedy this, recent economists have often extended the theory of NM utility, and the subsequent theory of subjective expected utility discussed in Savage (1954) and Anscombe and Aumann (1963). Schmeidler (1989) and Gilboa and Schmeidler (1989) are examples of such extensions. It should also be noted that the continuity assumption is stronger than Harsanyi's assumption. That is, Harsanyi only assumed the existence of the NM utility function, and to ensure this, we only need segmental continuity. Ensuring the continuity of the NM utility function may be strong in this context.

However, when one explains that the NM utility function coincides with Alt's utility function because these axioms are strong, we have to scratch our heads. Indeed, in order to derive the coincidence of the NM utility function and Alt's utility function, the NM utility function must first exist, and because Alt's utility function is usually continuous, the NM utility function must also be continuous. But, could this be the essence of the problem? For example, Theorem 3 does not hold if we consider a model in which independence does not hold, because the NM utility function does not exist in the first place. However, it is hard to believe that this leads to a better understanding of Theorem 3.

Therefore, our position is that although the existence of the NM utility function and the assumption for continuity are not necessarily weak, we cannot conclude that they are the main cause of the paradoxical conclusion of Theorem 3.

\subsection{Existence and Properties of Alt's system}
The next thing to be discussed is the existence of an Alt's system $\ge_i$ that matches the preference relation $\succsim_i$ and the axioms imposed on this system. The discussion in the previous subsection applies directly here. In deriving the coincidence between the NM utility function and Alt's utility function, Alt's utility function must exist, and thus Theorem 3 does not hold for a model in which Alt's utility function does not exist. However, this argument does not seem to provide a solution to our problem.

We now consider an additional issue. The existence of $\ge_0$ that matches $\succsim_0$ may seem to be strong. In the usual utilitarian argument, the change in utility level must express the intensity of improvement, which requires the existence of an Alt's system $\ge_i$ that matches $\succsim_i$ for each individual $i$. However, it is not usually assumed that the society has an Alt's system $\ge_0$. Thus, this may be a stronger assumption than the usual utilitarian argument.

In fact, the interpretation of $\succsim_0$ changes the evaluation of the strength of this assumption. We only described $\succsim_0$ as expressing some ethical order. However, in the explanation of Harsanyi (1955), who started to represent the society in this form, $\succsim_0$ is assumed to be possessed by individuals. There is a special individual $i$ in the society who is assumed to possess an ethical order $\succsim_0$ and a preference order $\succsim_i$. If we comply with this idea, the ethical order is also an evaluation criterion owned by some individual, and thus the existence of $\ge_0$ may be regarded as not a strong assumption.

\subsection{Evaluation of Harsanyi's and Harvey's Additional Axioms}
Finally, let us evaluate axiom (i) of Theorem 1 and axiom (I) of Theorem 2. Axiom (I) of Theorem 2 requires that if two improvements are equivalent for every individual, then these are also ethically equivalent. This requirement seems to be not strong.

Axiom (i), in contrast, becomes very problematic upon closer inspection. Consider a simple separable society where $X=\mathbb{R}^n_+$ and $x_i$ represents the money owned by $i$. The NM utility function $u_i$ is a function of the variable $x_i$. Then, various concepts that express the risk attitude, such as certainty equivalent and absolute/relative risk aversion, can be defined. Axiom (i) then implies that the ethical risk attitude and the individual risk attitude must be exactly the same. Indeed, consider lotteries $P$ and $Q$ that only affect $x_i$. Whatever these lotteries may be, the meaning of axiom (i) is that the ethical evaluation between $P$ and $Q$ must be perfectly consistent with the evaluation by individual $i$. In particular, any measures of risk aversions for the ethical preference must be the same as that for the individual preference.

Responding to criticisms by Diamond (1967) and Sen (1970) for imposing independence on ethical preference, Harsanyi (1975) stated that: ``Surely, when we act on behalf of other people, let alone when we act on behalf of society as a whole, we are under art obligation to follow, if anything, higher standards of rationality than when we are dealing with our own private affairs.'' Sen (1976) stated that he does not believe that a ``higher standard of rationality'' means the independence axiom. Our thinking is similar to Sen's opinion. We believe that a ``higher standard of rationality'' usually refers to ``higher risk aversion''. In our view, when one works on behalf of society, he/she will act much more ``carefully'' than when he/she just minds his/her personal business, and this carefulness should surface as a risk-averse attitude. From this perspective, Harsanyi's axiom (i) rather seems to negate a ``higher standard of rationality'' in that it prohibits society from having higher risk aversion than individuals.

In conclusion, we have determined that axiom (i) in Theorem 1 is the only strong assumption in Theorem 3. For those who can accept this axiom, there is no difference between Alt's utility function and the NM utility function. Conversely, those who support the Luce-Raiffa type argument that Alt's utility function is different from the NM utility function must reject this axiom.

\subsection{Another Possibility of Society}
We have so far taken what we have named ``society'' to correspond to the actual society. However, since our definition of ``society'' is only mathematical, it could be transferred to a completely different problem. Suppose, for example, that you are thinking about today's dinner. Good food is obviously welcome, but you have been warned by a doctor about high blood pressure and high blood sugar, and you do not want to eat too much salt or sugar. Thus, you have multiple criteria even in a very personal decision-making problem, and you must take all these criteria into account. In this case, if $\succsim_1$ is a preference for subjective taste evaluation, $\succsim_2$ is a preference for the amount of salt, and $\succsim_3$ is a preference for the amount of sugar, then the overall preference $\succsim_0$ is derived from the triplet $(\succsim_1,\succsim_2,\succsim_3)$. This has the same mathematical structure as our ``society''.

In this case, all the problems mentioned in the previous section for axiom (i) vanish, because we switch the problem to comparing risk attitudes within the same individual. Therefore, if all other axioms are unproblematic, then the conclusion of Theorem 3 holds, and the individual's NM utility function and Alt's utility function are identical. However, even in this case, some of the axioms supporting Theorem 3, including semi-separability, will have to be scrutinized more rigorously.

\section{Conclusion}
We derived two variants of the known utilitarian theorems; one is for NM utility functions, and another is for Alt's utility functions. Moreover, we derived that, under the axioms of these theorems, NM and Alt's utility functions must coincide.

From an optimistic viewpoint, this result is useful for applied research because it allows the estimation of Alt's utilities, which is beneficial for welfare analysis because of its interpretation, through the estimation of NM utilities. From a pessimistic viewpoint, however, it may indicate at least one of the assumptions of utilitarian theorems is incredibly strong. Upon closer inspection, we found that (i) of the axiom in Theorem 1 is strong, and the rest is not strong.

Hence, we succeeded in modernizing Luce--Raiffa's critique. That is, if we should not be identified as NM utility and Alt's utility, then axiom (i) must be denied. Conversely, if one thinks that axiom (i) is not strong, then Alt's utility must coincide with NM utility.

\appendix
\section{Proofs}
\subsection{Proof of Theorem 1}
It is clear that (ii) implies (i), and thus we only show that (i) implies (ii).

First, we show a lemma on linear algebra.

\vspace{12pt}
\noindent
{\bf Lemma 4}. Suppose that $V$ is a vector space, $N\ge 1$, and $f_i:V\to \mathbb{R}$ for $i\in \{0,...,N\}$ are linear functionals. Then, the following two statements are equivalent.
\begin{enumerate}[1)]
\item There exist $a_1,...,a_N\in\mathbb{R}$ such that $f_0=\sum_{i=1}^Na_if_i$.

\item $f_1(v)=...=f_N(v)=0$ implies that $f_0(v)=0$.
\end{enumerate}

\vspace{12pt}
\noindent
{\bf Proof of Lemma 4}. It is clear that 1) implies 2), and thus we only show the converse. We use mathematical induction. First, suppose $N=1$ and 2) holds. If $f_1\equiv 0$, then $f_0\equiv 0$, and if we define $a_1=1$, 1) holds. Hence, we assume that there exists $v\in V$ such that $f_1(v)\neq 0$. Define $a_1=\frac{f_0(v)}{f_1(v)}$. Choose any $w\in V$. For $b=\frac{f_1(w)}{f_1(v)}$,
\[f_1(w-bv)=f_1(w)-bf_1(v)=0,\]
and by 2), we have that $f_0(w-bv)=0$, which implies that $f_0(w)=bf_0(v)$. Hence,
\[f_0(w)=bf_0(v)=\frac{f_1(w)}{f_1(v)}a_1f_1(v)=a_1f_1(w),\]
and thus 1) holds.

Next, suppose that this theorem holds for $N=k$, and consider the case in which $N=k+1$. Suppose that 2) holds. If $f_1(v)=...=f_k(v)=0$ implies that $f_0(v)=0$, then by the induction hypothesis, there exist $a_1,...,a_k$ such that $f_0=\sum_{i=1}^ka_if_i$. Hence, if we define $a_{k+1}=0$, then 1) holds. Therefore, we assume that there exists $v\in V$ such that
\[f_1(v)=...=f_k(v)=0,\ f_0(v)\neq 0.\]
By 2), we have that $f_{k+1}(v)\neq 0$. Define $a_{k+1}=\frac{f_0(v)}{f_{k+1}(v)}$ and $g=f_0-a_{k+1}f_{k+1}$. Suppose that $f_1(w)=...=f_k(w)=0$. For $b=\frac{f_{k+1}(w)}{f_{k+1}(v)}$,
\[f_{k+1}(w-bv)=f_{k+1}(w)-bf_{k+1}(v)=0,\]
and by definition, $f_i(v)=f_i(w)=0$ for any $i\in \{1,...,k\}$, which implies that $f_i(w-bv)=0$. Therefore, by 2), we have that $f_0(w-bv)=0$, and thus $f_0(w)=bf_0(v)$. Hence,
\[f_0(w)=bf_0(v)=\frac{f_{k+1}(w)}{f_{k+1}(v)}a_{k+1}f_{k+1}(v)=a_{k+1}f_{k+1}(w).\]
Thus, we have that $g(w)=0$. To summarize the above arguments, we have found that $f_1(w)=...=f_k(w)=0$ implies that $g(w)=0$. Hence, by the induction hypothesis, there exist $a_1,...,a_k$ such that
\[g=\sum_{i=1}^ka_if_i,\]
which implies that 1) holds. This completes the proof. $\blacksquare$

\vspace{12pt}
For the convenience of later arguments, we define $u_0(x)\equiv 1$. We separate the proof into three steps. 

\vspace{12pt}
\noindent
{\bf Step 1}. If $X$ is a finite set $\{x_1,...,x_m\}$, then (ii) holds. Moreover, if $(u_i)_{i=0}^n$ is linearly independent, then $a_1,...,a_n$ and $b$ are uniquely determined.

\vspace{12pt}
\noindent
{\bf Proof of Step 1}. Choose any $\eta\in \mathbb{R}^m$ such that
\[\sum_{j=1}^m\eta_ju_i(x_j)=0\]
for all $i\in \{0,...,n\}$, and choose $\lambda>0$ so small that if we define
\[P(\{x_j\})=\frac{1}{m}+\lambda\eta_j,\ Q(\{x_j\})=\frac{1}{m}-\lambda\eta_j,\]
then $P,Q\in \mathscr{P}$. By assumption, $E_P[u_i]=E_Q[u_i]$ for all $i\in \{1,...,n\}$. Therefore, by (i), we have that $E_P[v]=E_Q[v]$. This implies that
\[\sum_{j=1}^m\eta_jv(x_j)=0.\]
Define an $(n+1)\times m$ matrix $A$ whose $(i,j)$-th element is $u_{i-1}(x_j)$, and a vector $\beta$ whose $j$-th element is $v(x_j)$. Then, we have shown that
\[A\eta=0\Rightarrow \beta^T\eta=0.\]
By Lemma 3, there exist $b\in\mathbb{R}$ and $a_1,...,a_n\in \mathbb{R}$ such that
\[\beta^T=(b,a_1,...,a_n)A.\]
Let $e_j$ be the $j$-th unit vector. Then,
\[v(x_j)=\beta^Te_j=(b,a_1,...,a_n)Ae_j=b+\sum_{i=1}^na_iu_i(x_j),\]
as desired.

If $(u_i)_{i=0}^n$ is linearly independent, then $A$ has an $(n+1)\times (n+1)$ regular submatrix $B$. By the regularity of $B$, $(b,a_1,...,a_n)$ must be unique. This completes the proof of Step 1. $\blacksquare$

\vspace{12pt}
\noindent
{\bf Step 2}. Suppose that $X$ is a finite set $\{x_1,...,x_m\}$, and $(u_i)_{i=0}^n$ is linearly independent. Moreover, suppose that the society obeys the Pareto criterion. Then, $a_i>0$ for all $i\in \{1,...,n\}$.

\vspace{12pt}
\noindent
{\bf Proof of Step 2}. Let $A$ be the $(n+1)\times m$ matrix whose $(i,j)$-th element is $u_{i-1}(x_j)$. Because $(u_i)_{i=0}^n$ is linearly independent, there exists an $(n+1)\times (n+1)$ regular submatrix $B$ of $A$. Without loss of generality, we assume that the $(i,j)$-th element of $B$ is $u_{i-1}(x_j)$. We only show that $a_1>0$, because the rest cases can be treated in the same manner. Choose $\eta\in \mathbb{R}^m$ such that $\eta_j=0$ if $j\ge n+2$ and
\[(\eta_1,...,\eta_{n+1})^T=B^{-1}(1,1,0,...,0)^T,\]
and choose any sufficiently small $\lambda>0$ such that if
\[P(\{x_j\})=\frac{1}{m}+\lambda \eta_j,\ Q(\{x_j\})=\frac{1}{m}-\lambda \eta_j,\]
Then $P,Q\in\mathscr{P}$. Then, $E_P[u_1]>E_Q[u_1]$ and $E_P[u_i]=E_Q[u_i]$ for all $i\in \{2,...,n\}$, and by the Pareto criterion, we have that $E_P[v]>E_Q[v]$. However,
\[E_P[v]-E_Q[v]=\sum_{i=1}^na_i(E_P[u_i]-E_Q[u_i])=a_1(E_P[u_1]-E_Q[u_1]),\]
which implies that $a_1>0$. This completes the proof of Step 2. $\blacksquare$

\vspace{12pt}
\noindent
{\bf Step 3}. Condition (ii) holds for an arbitrary $X$. In addition, the society obeys the Pareto criterion if and only if we can choose $a_1,...,a_n,b$ so that $a_i>0$ for all $i\in \{1,...,n\}$.

\vspace{12pt}
\noindent
{\bf Proof of Step 3}. If $u_i$ is a constant function for every $i\in \{1,...,n\}$, then by (i), we have that $v$ is also a constant function, and thus our claim is trivially correct. Hence, we assume that there exists $i\in \{1,...,n\}$ such that $u_i$ is not a constant function. Choose $M\subset \{1,...,n\}$ such that for any $j\in N$, there uniquely exists $(c_i^j)_{i\in M\cup \{0\}}$ such that $u_j=c_0^ju_0+\sum_{i\in M}c_i^ju_i$. By easy inductive arguments, we can show the existence of such a set $M$. In this case, $(u_i)_{i\in M\cup \{0\}}$ is linearly independent. Renumbering if necessary, we can assume that $M=\{1,...,m\}$. By the construction of $M$, there exist $x_0,...,x_m\in X$ such that a matrix $A$ whose $(i,j)$-th element is $u_{i-1}(x_{j-1})$ is regular. Note that, for any $P,Q\in\mathscr{P}$, if $E_P[u_i]=E_Q[u_i]$ for all $i\in M$, then $E_P[u_i]=E_Q[u_i]$ for all $i\in \{1,...,n\}$, and thus $E_P[v]=E_Q[v]$. Let $Y=\{x_0,...,x_m\}$. By Step 1, there uniquely exist $a_1,...,a_m\in \mathbb{R}$ and $b\in \mathbb{R}$ such that
\[v(x)=\sum_{i=1}^ma_iu_i(x)+b\]
for every $x\in Y$. Choose any $x^*\in X\setminus Y$, and define $Z=Y\cup \{x^*\}$. This is also a finite set, and thus Step 1 implies that there exist $a_1',...,a_m'\in \mathbb{R}$ and $b'\in\mathbb{R}$ such that
\[v(x)=\sum_{i=1}^ma_i'u_i(x)+b'\]
for every $x\in Z$. However, because $a_i$ and $b$ are unique, we have that $a_i'=a_i$ and $b'=b$. In particular,
\[v(x^*)=\sum_{i=1}^ma_iu_i(x^*)+b.\]
Because $x^*$ is arbitrary, we have that
\[v(x)=\sum_{i=1}^ma_iu_i(x)+b\]
for every $x\in X$, and thus if we define $a_i=0$ for $i\in \{m+1,...,n\}$, then (ii) holds.

Now, suppose that the society obeys the Pareto criterion. Then, in the above arguments, $a_i>0$ for all $i\in \{1,...,m\}$. If $m=n$, then our claim holds. Hence, suppose that $m<n$. Choose $\varepsilon>0$ such that for all $i\in \{1,...,n\}$, $\sum_{j=m+1}^n\varepsilon|c_i^j|<a_i$. Define $a_i^*=\varepsilon$ for $i\in \{m+1,...,n\}$, and
\[b^*=b-\sum_{j=m+1}^na_j^*c_0^j,\]
\[a_i^*=a_i-\sum_{j=m+1}^na_j^*c_i^j\]
for any $i\in \{1,...,m\}$. Then, $a_i^*>0$ for all $i\in \{1,...,n\}$, and
\begin{align*}
v=&~\sum_{i=1}^ma_iu_i+b\\
=&~\sum_{i=1}^ma_i^*u_i+b^*+\sum_{i=1}^m(a_i-a_i^*)u_i+(b-b^*)\\
=&~\sum_{i=1}^ma_i^*u_i+b^*+\sum_{i=0}^m\sum_{j=m+1}^na_j^*c_i^ju_i\\
=&~\sum_{i=1}^ma_i^*u_i+b^*+\sum_{j=m+1}^na_j^*u_j\\
=&~\sum_{i=1}^na_i^*u_i+b^*,
\end{align*}
as desired.

Conversely, if (ii) holds and $a_i>0$ for all $i\in \{1,...,n\}$, then clearly the society obeys the Pareto criterion. This completes the proof of Step 3. $\blacksquare$

\vspace{12pt}
Step 3 states that Theorem 1 is correct. This completes the proof. $\blacksquare$

\subsection{Proof of Theorem 2}
It is clear that (II) implies (I), and thus we only show that (I) implies (II).

Define a vector-valued function $u(x)=(u_1(x),...,u_n(x))$. Let $I_i=u_i(X)$. Because of the continuity of $u_i$ and the connectedness of $X$, $I_i$ is either a singleton or an interval.\footnote{In this paper, we call a subset $I$ of $\mathbb{R}$ an {\bf interval} if it is convex and contains at least two numbers.} Define $I=\prod_{i=1}^nI_i$. Because the society is semi-separable, we have that $u(X)=I$.

Define $U_i(x,y)=u_i(x)-u_i(y)$, $J_i=U_i(X^2)$, and $J=\prod_{i=1}^nJ_i$. Moreover, define $U(x,y)=u(x)-u(y)$ and $V(x,y)=v(x)-v(y)$. By definition, $J_i$ is either $\{0\}$ or an interval including $0$. Because the society is semi-separable, for any $c\in J$, there exists $(x,y)\in X^2$ such that $c=U(x,y)$. If $c=U(z,w)$ for $(z,w)\in X^2$, then $[x,y]=_i[z,w]$ for any $i\in \{1,...,n\}$, and thus by (I), we have that $[x,y]=_0[z,w]$. This implies that $V(x,y)=V(z,w)$. Thus, we can define $F(c)=V(x,y)$. This $F$ is a well-defined real-valued function on $J$, and
\[V(x,y)=F(U(x,y))\]
for any $(x,y)\in X^2$.

Next, we show that for any $c,c',c''\in I$,
\begin{equation}
F(c'-c)+F(c''-c')=F(c''-c).
\end{equation}
Because $I=u(X)$, there exist $x,y,z\in X$ such that $u(x)=c, u(y)=c', u(z)=c''$. Then, $c'-c=U(y,x)$, $c''-c'=U(z,y)$, and $c''-c=U(z,x)$, and thus,
\begin{align*}
F(c'-c)+F(c''-c')=&F(U(y,x))+F(U(z,y))\\
=&~V(y,x)+V(z,y)=v(y)-v(x)+v(z)-v(y)\\
=&~v(z)-v(x)=V(z,x)=F(U(z,x))=F(c''-c),
\end{align*}
as desired. Hence, (3) is correct.

Let $e_i$ be the $i$-th unit vector, and define $F_i(c)=F(ce_i)$ for any $c\in J_i$. We can easily show that
\[F(c)=\sum_{i=1}^nF_i(c_i).\]
Obviously, if $c,c',c''\in I_i$, then
\begin{equation}
F_i(c'-c)+F_i(c''-c')=F_i(c''-c).
\end{equation}
We will show that if $c,c',c+c'\in J_i$, then
\begin{equation}
F_i(c)+F_i(c')=F_i(c+c').
\end{equation}
Suppose that $c,c'\ge 0$. Then, there exist $x,z\in X$ such that $u_i(z)-u_i(x)=c+c'$. Because $I_i$ is an interval, there exist $y\in X$ such that $u_i(y)-u_i(x)=c$. Therefore, by (4),
\begin{align*}
F_i(c)+F_i(c')=&~F_i(u_i(y)-u_i(x))+F_i(u_i(z)-u_i(y))\\
=&~F_i(u_i(z)-u_i(x))=F_i(c+c'),
\end{align*}
as desired. Hence, (5) holds in this case. By the symmetrical arguments, we can show that if $c,c'\le 0$, then (5) holds. Next, suppose that $c<0<c'$. Choose $x,w\in X$ such that $u_i(w)-u_i(x)\ge \max\{-c,c'\}$. Because $I_i$ is an interval, there exist $y,z\in X$ such that $u_i(y)-u_i(x)=-c$ and $u_i(z)-u_i(x)=c'$. Therefore, by (4),
\begin{align*}
F_i(c)+F_i(c')=&~F_i(u_i(x)-u_i(y))+F_i(u_i(z)-u_i(x))\\
=&~F_i(u_i(z)-u_i(y))=F_i(c+c'),
\end{align*}
as desired. Hence, (5) holds in all cases.

Because $0\in J_i$, this implies that $F_i(0)=2F_i(0)$, and thus $F_i(0)=0$. Moreover, if $c\in J_i$, then $-c\in J_i$, and by (5), $0=F_i(0)=F_i(c)+F_i(-c)$, which implies that $F_i(-c)=-F_i(c)$. If $2c\in J_i$, then $F_i(2c)=2F_i(c)$. Hence, $F_i(c)/2=F_i(c/2)$. Applying mathematical induction to the above results, we can show that for every dyadic rationals $x_{k,\ell}=\frac{\ell}{2^k}$, if $c\in J_i$ and $x_{k,\ell}c\in J_i$, then $F_i(x_{k,\ell}c)=x_{k,\ell}F_i(c)$.

Suppose that $c\in J_i$ and $c>0$. Then, there exist $x,y\in X$ such that $U_i(x,y)=c$ and $U_j(x,y)=0$ for all $j\in \{1,...,n\}\setminus \{i\}$. Because of the Pareto criterion, we have that $V(x,y)>0$. Therefore,
\[F_i(c)=\sum_{j=1}^nF_j(U_j(x,y))=F(U(x,y))=V(x,y)>0.\]
By (5), $F_i$ is increasing.

Suppose that $x\in \mathbb{R}$ and $c,xc\in J_i$. If $c=0$, then $F_i(xc)=xF_i(c)=0$. If $c>0$, then let $\ell(k)$ be the unique $\ell$ such that $x_{k,\ell}\le x<x_{k,\ell+1}$. If $xc$ is not an endpoint of $J_i$, then $x_{k,\ell(k)}c,x_{k,\ell(k)+1}c\in J_i$ for any sufficiently large $k$. Because $F_i$ is increasing,
\[x_{k,\ell(k)}F_i(c)=F_i(x_{k,\ell(k)}c)\le F_i(xc)\le F_i(x_{k,\ell(k)+1}c)=x_{k,\ell(k)+1}F_i(c),\]
and letting $k\to \infty$, we have that
\[F_i(xc)=xF_i(c).\]
In the general case, we have that $\frac{xc}{2}$ is not an endpoint of $J_i$. Therefore,
\[F_i(xc)=2F_i(xc/2)=xF_i(c),\]
as desired. We can easily show that the same result holds when $c<0$. Therefore, $F_i$ is linear, and thus there exists $a_i>0$ such that $F_i(c)=a_ic$.

Hence, $F(c)=\sum_{i=1}^na_ic_i$. Fix an $x^*\in X$. Then,
\[v(x)-v(x^*)=V(x,x^*)=F(U(x,x^*))=\sum_{i=1}^na_i(u_i(x)-u_i(x^*))\]
and thus, if we define
\[b=v(x^*)-\sum_{i=1}^na_iu_i(x^*),\]
then (II) holds. This completes the proof. $\blacksquare$

\subsection{Proof of Lemma 1}
Suppose that $Y$ is the set of all bounded and continuous real-valued functions on $X$. Then, $Y$ is a Banach space with respect to the sup norm. For $P\in \mathscr{P}$, define a function $T^*(P)$ as $T^*(P)(u)=E_P[u]$. Then, $T^*(P)\in Y'$. Suppose that $P,Q\in \mathscr{P}$ and $P\neq Q$. Then, there exists $x\in X$ such that $P(\{x\})\neq Q(\{x\})$. Let $A=\{y\in X|\max\{P(\{y\}),Q(\{y\})\}>0\}$. Then, $A$ is a finite set. Because $X$ is Tychonoff, it is Hausdorff, and thus $A$ is closed. Hence, there exists a continuous function $f:X\to [0,1]$ such that $f(x)=1$ and $f(y)=0$ for all $y\in A$. Then, $f\in Y$ and
\[T^*(P)(f)=P(\{x\})\neq Q(\{x\})=T^*(Q)(f),\]
which implies that the mapping $T^*:\mathscr{P}\to Y'$ is one-to-one, and thus we can treat $\mathscr{P}$ as the subset of $Y'$.

For any $u\in Y$, let $L_u:Y'\to \mathbb{R}$ be defined as $L_u(T)=T(u)$. Then, $L_u\in Y''$. Let $\sigma_1$ be the set of all $L_u^{-1}(A)$, where $u\in Y$ and $A$ is an open set in $\mathbb{R}$, $\sigma_2$ be the set of all finite intersections of sets in $\sigma_1$, and $\tau$ be the set of all unions of families in $\sigma_2$. Then, it is easy to show that $\tau$ is the least topology such that $L_u$ is continuous for all $u\in Y$. Because $T^*$ is one-to-one, the relative topology on $\mathscr{P}$ with respect to $\tau$ is Hausdorff.

Now, choose any net $(P_{\nu})$ in $\mathscr{P}$ and $P\in \mathscr{P}$. Suppose that $(P_{\nu})$ converges to $P$ in the topology $\tau$. Choose any $u\in Y$. Because $L_u$ is continuous under $\tau$, we have that the net $(E_{P_{\nu}}[u])$ converges to $E_P[u]$. Conversely, suppose that $(E_{P_{\nu}}[u])$ converges to $E_P[u]$ for all $u\in Y$. Choose any neighborhood $U$ of $P$. By construction, there exists $V\in \sigma_2$ such that $P\in V$ and $V\subset U$. Because $V\in \sigma_2$, there exist $V_1,...,V_m\in \sigma_1$ such that $V=V_1\cap...\cap V_m$. Suppose that $V_i=L_{u_i}^{-1}(A_i)$, where $u_i\in Y$ and $A_i$ is an open set in $\mathbb{R}$. Because $P\in V_i$, $E_P[u_i]\in A_i$. Because $(E_{P_{\nu}}[u_i])$ converges to $E_P[u_i]$, there exists $\nu_i$ such that if $\nu\ge \nu_i$, then $E_{P_{\nu}}[u_i]\in A_i$. By the definition of the directed set, there exists $\nu^*$ such that $\nu^*\ge \nu_i$ for all $i\in \{1,...,m\}$. If $\nu\ge \nu^*$, then $E_{P_{\nu}}[u_i]\in A_i$ for all $i\in \{1,...,m\}$, which implies that $P_{\nu}\in V\subset U$. Therefore, $(P_{\nu})$ converges to $P$ in the topology $\tau$. This completes the proof of Lemma 1. $\blacksquare$

\subsection{Proof of Lemma 2}
First, suppose that $\succsim$ is sequentially continuous. Let $P\succ Q\succ R$. Choose $n$ such that $X_n$ includes the supports of $P,Q,R$, and let $(t_k)$ be a sequence on $[0,1]$ such that $(1-t_k)P+t_kR\succsim Q$ for any $k$ and $t_k\to t^*$ as $k\to \infty$. Then, $(1-t_k)P+t_kR\succsim_{X_n}Q$. Moreover, $(1-t_k)P+t_kR\to (1-t^*)P+t^*R$ in the weak* topology, and thus $(1-t^*)P+t^*R\succsim_{X_n}Q$, which implies that $(1-t^*)P+t^*R\succsim Q$. Hence, $\{t\in [0,1]|(1-t)P+tR\succsim Q\}$ is closed, as desired. By the symmetrical argument, we can show that $\{t\in [0,1]|Q\succsim (1-t)P+tR\}$ is closed, and thus $\succsim$ is segmentally continuous.

Second, suppose that $\succsim$ is independent and closed with respect to the weak* topology, and let $u:X\to \mathbb{R}$ be an NM utility function of $\succsim$.\footnote{In this case, $\succsim$ is sequentially continuous with $X_n\equiv X$.} Suppose that $u$ is not upper semi-continuous at $x$. Then, there exists $\varepsilon>0$ such that for any neighborhood $V$ of $x$, there exists $x_V\in V$ such that $u(x_V)-u(x)\ge \varepsilon$. Then, $(x_V)$ consists of a net that converges to $x$, and by the definition of the weak* topology, $(\delta_{x_V})$ converges to $\delta_x$. Choose any $V^*$ and let $P=(1-t)\delta_x+t\delta_{x_{V^*}}$, where $t\in ]0,1[$ is sufficiently small that $0<E_P[u]-u(x)<\varepsilon$. Then, $\delta_{x_V}\succeq P$ for all $V$, but $\delta_x\not\succeq P$, which is a contradiction. Therefore, $u$ is upper semi-continuous. By the symmetrical arguments, we can show that $u$ is lower semi-continuous, and thus it is continuous. Next, suppose that $u$ is unbounded from above. Choose any $x^*,x_0\in X$ such that $u(x_0)>u(x^*)$, and for any $n\ge 1$, choose $x_n\in X$ such that $u(x_n)-u(x^*)\ge 2^n(u(x_0)-u(x^*))$. Define
\[P_n=\frac{1}{2^n}\delta_{x_n}+\left(1-\frac{1}{2^n}\right)\delta_{x^*}.\]
Then, $P_n\to \delta_{x^*}$ with respect to the weak* topology. On the other hand,
\[E_{P_n}[u]\ge E_{\delta_{x_0}}[u]>E_{\delta_{x^*}}[u]\]
which implies that $\succsim$ is not closed, a contradiction. Therefore, we have that $u$ is bounded from above. By the symmetrical arguments, we can show that $u$ is bounded from below. Hence, if $\succsim$ is independent and closed with respect to the weak* topology, then $u$ is continuous and bounded.

Now, suppose that $\succsim$ is independent and sequentially continuous. Because $\succsim$ is sequentially continuous, it is segmentally continuous, and thus there exists an NM utility function $u:X\to \mathbb{R}$. Choose a sequence $(X_n)$ as in the definition of sequential continuity, and let $u_n$ be the restriction of $u$ to $X_n$. Then, $u_n$ is an NM utility function of $\succsim_{X_n}$. Because $\succsim_{X_n}$ is closed, we have that $u_n$ is continuous and bounded. Choose any $x\in X$. Then, there exists $n$ such that $x\in X_n$, and $X_n$ is included in the interior of $X_{n+1}$. Therefore, there exists a neighborhood $U$ of $x$ such that $U\subset X_{n+1}$. Because $u(y)=u_{n+1}(y)$ for all $y\in U$, $u$ is continuous at $x$. Since $x$ is arbitrary, we have that $u$ is continuous.

Third, suppose that $\succsim$ has a continuous and bounded NM utility function $u:X\to \mathbb{R}$. Choose any net $(P_{\nu},Q_{\nu})$ on $\succsim$ that converges to $(P,Q)\in \mathscr{P}^2$ with respect to the weak* topology. Because $u$ is continuous and bounded, we have that the net $(E_{P_{\nu}}[u],E_{Q_{\nu}}[u])$ converges to $(E_P[u],E_Q[u])$. This implies that $E_P[u]\ge E_Q[u]$, and thus $P\succsim Q$. Therefore, $\succsim$ is closed with respect to the weak* topology.

Now, suppose that $\succsim$ has a continuous NM utility function $u:X\to \mathbb{R}$. Define
\[X_n=\{x\in X||u(x)|\le n\}.\]
Let
\[U_n=u^{-1}(]-n-1/2,n+1/2[).\]
Because $u$ is continuous, $X_n$ is closed and $U_n$ is open in $X$. Moreover, $X_n\subset U_n\subset X_{n+1}$. Therefore, we conclude that $X_n$ is included in the interior of $X_{n+1}$. Let $u_n$ be the restriction of $u$ to $X_n$. Then, $u_n$ is a continuous and bounded NM utility function of $\succsim_{X_n}$, which implies that $\succsim_{X_n}$ is closed with respect to the weak* topology. This completes the proof. $\blacksquare$

\subsection{Proof of Lemma 3}
Suppose that $u:X\to \mathbb{R}$ is a continuous Alt's utility function of $\ge$, and $v:X\to \mathbb{R}$ is another Alt's utility function. If $u$ is a constant function, then $v$ is also a constant function, and thus $v$ is a positive affine transform of $u$. Therefore, we assume that $u$ is not a constant function. Let $I=u(X)$. Because $u$ is continuous and $X$ is connected, we have that $I$ is an interval. Define $\succsim=\{(x,y)\in X^2|u(x)\ge u(y)\}$. Then, $\succsim$ matches $\ge$, and thus $x\succsim y$ if and only if $v(x)\ge v(y)$.

Now, choose $x,y\in X$ such that $x\succ y$. Without loss of generality, we can assume that $u(x)=1,u(y)=0$. For each dyadic rational $r=\frac{i}{2^m}$, if $r\in I$, then there exists $z_r\in X$ such that $u(z_r)=r$. If $r=1/2$, then
\[u(x)-u(z^{1/2})=u(z^{1/2})-u(y)=\frac{1}{2},\]
and thus $[z^1,z^{1/2}]=[z^{1/2},z^0]$. Therefore,
\[v(x)-v(z^{1/2})=v(z^{1/2})-v(y),\]
which implies that
\[v(z^{1/2})-v(y)=\frac{1}{2}(v(x)-v(y)).\]
Inductively, we can show that for all $m$,
\[v(z^{1/2^m})-v(y)=\frac{1}{2^m}(v(x)-v(y)).\]
Moreover, if $r=\frac{i}{2^m}$ and $r'=\frac{i+1}{r^m}$, then
\[u(z^{r'})-u(z^r)=u(z^{1/2^m})-u(y),\]
which implies that $[z^{r'},z^r]=[z^{1/2^m},y]$. Thus,
\[v(z^{r'})-v(z^r)=v(z^{1/2^m})-v(y)=\frac{1}{2^m}(v(x)-v(y)),\]
and hence, for any dyadic rational $r\in I$,
\[v(z^r)-v(y)=r(v(x)-v(y)).\]
Now, choose any $z\in X$. If $u(z)$ is not the minimum value of $u$, then for sufficiently large $m$, there exists $i_m$ such that if $r=\frac{i_m}{2^m},r'=\frac{i_m+1}{2^m}$, then $r\in I$ and $r\le u(z)\le r'$. Then, $[z^{1/2^m},y]\ge [z,z^r]$, and thus,
\[0\le \frac{v(z)-v(z^r)}{v(x)-v(y)}\le \frac{1}{2^m}.\]
Hence,
\[\frac{v(z)-v(y)}{v(x)-v(y)}=\frac{v(z)-v(z^r)}{v(x)-v(y)}+\frac{v(z^r)-v(y)}{v(x)-v(y)}\in [r,r'],\]
and letting $m\to \infty$, we have that
\[\frac{v(z)-v(y)}{v(x)-v(y)}=u(z),\]
which is equivalent to
\[v(z)=(v(x)-v(y))u(z)+v(y),\]
as desired. The case in which $u(z)$ is the minimum value of $u$ can be treated similarly. This completes the proof. $\blacksquare$

\subsection{Proof of Theorem 3}
First, we prove the following proposition.

\vspace{12pt}
\noindent
{\bf Proposition 1}. Suppose that $X$ is a connected Hausdorff topological space, and $(u_1,...,u_n)$ and $(u_1^*,...,u_n^*)$ are two profiles of continuous real-valued functions on $X$. Let $v=\sum_{i=1}^nu_i$ and $v^*=\sum_{i=1}^nu_i^*$. Suppose also that $v,v^*$ represent the same weak order $\succsim_0$, and for each $i\in \{1,...,n\}$, $u_i,u_i^*$ represent the same weak order $\succsim_i$. Let $I_i=u_i(X)$, $u=(u_1,...,u_n)$, and $I=u(X)$. If $I=\prod_{i=1}^nI_i$ and there are $j^+,k^+\in \{1,...,n\}$ such that $u_{j^+},u_{k^+}$ are not constant, then $u_i^*$ is a positive affine transformation of $u_i$ for all $i\in \{1,...,n\}$.

\vspace{12pt}
\noindent
{\bf Proof of Proposition 1}. Note that, because $X$ is connected and $u_i$ is continuous, $I_i$ is either a singleton or an interval.

Choose any $i\in \{1,...,n\}$. If $I_i$ is a singleton, then both $u_i^*$ and $u_i$ are constant functions. Thus, in this case, our claim is correct. Therefore, we assume that $I_i$ is an interval. By assumption, there exists $j\neq i$ such that $I_j$ is also an interval. Because $I=\prod_{k=1}^nI_k$, we have that there exist $x_i,y_i,x_j,y_j\in X$ such that
\[u_i(x_i)>u_i(y_i)=u_i(x_j)=u_i(y_j),\]
\[u_j(x_j)>u_j(y_j)=u_j(x_i)=u_j(y_i),\]
\[u_k(x_i)=u_k(y_i)=u_k(x_j)=u_k(y_j)\]
if $i\neq k\neq j$, and
\[u_i(x_i)-u_i(y_i)=u_j(x_j)-u_j(y_j)>0.\]
Replacing $u_k$ with $w_k(x)=(u_i(x_i)-u_i(y_i))^{-1}(u_k(x)-u_k(y_i))$, we can assume that $u_i(x_i)=u_j(x_j)=1,u_i(y_i)=u_j(y_j)=0$, and for any $k\in \{1,...,n\}\setminus \{i,j\}$, $u_k(x_i)=0$. Define
\[\hat{X}=\{x\in X|u_k(x)=0\mbox{ if }k\neq i\}.\]
Because $I=\prod_{k=1}^nI_k$, we have that $u_i(\hat{X})=I_i$. Let $z^1=x_i,z^0=y_i$. Let $D_m$ be the set of all dyadic rationals $s=\frac{\ell}{2^m}$ such that $s\in I_i$, and define $D=\cup_mD_m$. For any $s\in D$, there exists $z^s\in \hat{X}$ such that $u_i(z^s)=s$. Define
\[\bar{X}=\{x\in X|u_k(x)=0\mbox{ if }i\neq k\neq j\}.\]
Because $I=\prod_{k=1}^nI_k$, we have that $u_j(\bar{X})=I_j$. Choose any $s\in D_m$. Because $I_j$ is convex, there exists $w^s\in \bar{X}$ such that $u_j(w^s)=2^{-m}$ and $u_i(w^s)=s$. Suppose that $s,s'\in D_m$ and $s'-s=\frac{1}{2^m}$, and let $s_m=\frac{1}{2^m}$. By definition,
\[u_j(w^s)=u_j(w^{s_m}),\ u_j(z^{s'})=u_j(z^0)\]
and thus,
\[w^s\sim_jw^{s_m},\ z^{s'}\sim_jz^0,\]
which implies that
\[u_j^*(w^s)=u_j^*(w^{s_m}),\ u_j^*(z^{s'})=u_j^*(z^0).\]
Moreover, because $u_i(w^s)=u_i(z^s)$, $w^s\sim_iz^s$, and thus
\[u_i^*(w^s)=u_i^*(z^s).\]
Again by definition,
\[u_i(z^{s'})+u_j(z^{s'})=u_i(w^s)+u_j(w^s),\]
and if $i\neq k\neq j$, then $u_k(w^s)=u_k(z^{s'})=0$. Therefore, by definition,
\[v(z^{s'})=\sum_{p=1}^nu_p(z^{s'})=\sum_{p=1}^nu_p(w^s)=v(w^s),\]
which implies that $z^{s'}\sim_0w_s$, and thus
\[v^*(z^{s'})=v^*(w^s).\]
If $i\neq k\neq j$, then $u_k(z^{s'})=u_k(w^s)$, and thus $u_k^*(z^{s'})=u_k^*(w^s)$. Therefore, we have that
\[u_i^*(z^{s'})+u_j^*(z^{s'})=u_i^*(w^s)+u_j^*(w^s),\]
which implies that
\[u_i^*(z^{s'})-u_i^*(z^s)=u_i^*(z^{s'})-u_i^*(w^s)=u_j^*(w^s)-u_j^*(z^{s'})=u_j^*(w^{s_m})-u_j^*(z^0),\]
where the right-hand side is independent of the choice of $s\in D_m$.

Define
\[c^*=u_i^*(z^1)-u_i^*(z^0).\]
By the above consideration, we have that for any $s\in D$,
\[u_i^*(z^s)-u_i^*(z^0)=c^*s=c^*(u_i(z^s)-u_i(z^0)).\]
Therefore, if $x\sim_iz^s$ for some $s\in D$, then
\begin{equation}
u_i^*(x)-u_i^*(z_0)=c^*(u_i(x)-u_i(z^0)).
\end{equation}
Now, choose any $x\in X$. Suppose that $x$ is not a greatest element on $u_i^*$. Fix any $\varepsilon>0$. Because $u_i^*$ is continuous, there exists $y\in X$ such that $u_i^*(y)>u_i^*(x)$ and $u_i^*(y)-u_i^*(x)<\varepsilon$. Then, $u_i(y)>u_i(x)$, and thus there exists $s\in D$ such that $u_i(y)>u_i(z^s)>u_i(x)$. This implies that $u_i^*(y)>u_i^*(z^s)>u_i^*(x)$, and thus,
\begin{align*}
u_i^*(x)-u_i^*(z^0)=&~u_i^*(x)-u_i^*(z^s)+u_i^*(z^s)-u_i^*(z^0)\\
>&~-\varepsilon+u_i^*(z^s)-u_i^*(z^0)\\
=&~-\varepsilon+c^*(u_i(z^s)-u_i(z^0))\\
>&~-\varepsilon+c^*(u_i(x)-u_i(z^0)).
\end{align*}
Because $\varepsilon>0$ is arbitrary, we have that
\[u_i^*(x)-u_i^*(z^0)\ge c^*(u_i(x)-u_i(z^0)).\]
Symmetrically, we can show that if $x$ is not a least element on $u_i^*$, then
\[u_i^*(x)-u_i^*(z^0)\le c^*(u_i(x)-u_i(z^0)).\]
Therefore, (6) holds for every $x\in X$ that is neither a greatest nor a least element.

Finally, suppose that $x\in X$ is a greatest element on $u_i^*$. Then, $x$ is not a least element on $u_i^*$, and thus
\[u_i^*(x)-u_i^*(z^0)\le c^*(u_i(x)-u_i(z^0)).\]
Suppose that
\[u_i^*(x)-u_i^*(z^0)<c^*(u_i(x)-u_i(z^0)).\]
Because $I_i$ is convex, there exists $y\in X$ such that
\[u_i^*(x)-u_i^*(z^0)<c^*(u_i(y)-u_i(z^0))=u_i^*(y)-u_i^*(z_0),\]
which implies that $u_i^*(y)>u_i^*(x)$, a contradiction. Hence, we have that (6) holds for such an $x\in X$. By symmetrical arguments, we can show that (6) holds for any least element $x\in X$ on $u_i^*$. This completes the proof. $\blacksquare$

\vspace{12pt}
Now, we prove Theorem 3 using Proposition 1. By the assumption, there exists a continuous NM utility function $v^*$ of $\succeq_0$, and for each $i\in \{1,...,n\}$, there exists a continuous NM utility function $u_i^*$ of $\succeq_i$. Moreover, there exists a continuous Alt's utility function $v$ of $\ge_0$, and for each $i\in \{1,...,n\}$, there exists a continuous Alt's utility function $u_i$ of $\ge_i$. By Theorems 1 and 2, there exist $a_1^*,...,a_n^*\in \mathbb{R}$ and $b^*\in \mathbb{R}$ such that
\[v^*(x)=\sum_{i=1}^na_i^*u_i^*(x)+b^*,\]
for any $x\in X$, and there exist $a_1,...,a_n\in \mathbb{R}_{++}$ and $b\in \mathbb{R}$ such that
\[v(x)=\sum_{i=1}^na_iu_i(x)+b\]
for any $x\in X$. Because of the assumptions, both $v^*$ and $v$ represent $\succsim_0$, and for each $i\in \{1,...,n\}$, both $u_i^*$ and $u_i$ represent $\succsim_i$. Because $a_i>0$, $a_iu_i$ is also an Alt's utility function of $\ge_i$, and thus we can assume without loss of generality that $a_i=1$ for all $i\in \{1,...,n\}$.

If $u_i$ is constant, then $u_i^*$ is also constant, and thus we can assume that $a_i^*=1$. If $u_i$ is not constant, choose any $x,y$ such that $u_i(x)>u_i(y)$ and $u_j(x)=u_j(y)$ for any $j\neq i$. Then, $x\succ_iy$, and thus $u_i^*(x)>u_i^*(y)$. Moreover, if $j\neq i$, then $x\sim_jy$, and thus $u_j^*(x)=u_j^*(y)$. In this case, $v(x)>v(y)$, and thus $v^*(x)>v^*(y)$. Therefore, $a_i^*>0$. Because $a_i^*u_i^*$ is also an NM utility function, we can assume without loss of generality that $a_i^*=1$. Hence,
\[v^*(x)=\sum_{i=1}^nu_i^*(x)+b^*,\ v(x)=\sum_{i=1}^nu_i(x)+b\]
for all $x\in X$. Because $v^*-b^*$ is also an NM utility function and $v-b$ is also an Alt's utility function, we can assume $b^*=b=0$, and thus,
\[v^*(x)=\sum_{i=1}^nu_i^*(x),\ v(x)=\sum_{i=1}^nu_i(x)\]
for any $x\in X$. If $I_i=u_i(X)$, $u=(u_1,...,u_n)$, and $I=u(X)$, then because the society is semi-separable, we have that $I=\prod_{i=1}^nI_i$. Moreover, there exists at least two $i^*,j^*\in \{1,...,n\}$ such that both $u_{i^*},u_{j^*}$ are not constant. Therefore, we can apply Proposition 1 for $(u_1,...,u_n)$ and $(u_1^*,...,u_n^*)$, and thus $u_i^*$ is a positive affine transform of $u_i$, as desired. This completes the proof. $\blacksquare$

\section*{Acknowledgments}
The author is grateful to Mitsunobu Miyake for his helpful comments and suggestions in the 29th DC conference and 2023 Autumn Meeting of Japanese Economic Association.

\section*{References}

\begin{description}
\item{[1]} Alt, F. 1936. \"Uber die Messbarkeit des Nutzens. Zeitschrift f\"ur National\"okonomie 7, 161-169. Translated by Schach, S. 1971. On the Measurability of Utility. in: Chipman, J. S., Hurwicz, L., Richter, M. K., Sonnenschein, H. F. (Eds.) Preferences, Utility and Demand. Harcourt Brace Jovanovich, New York, pp.424-431.

\item{[2]} Anscombe, F. J., Aumann, R. J. 1963. A Definition of Subjective Probability. Ann. Math. Stat. 34, 199-205.

\item{[3]} Diamond, P. 1967. Cardinal Welfare, Individualistic Ethics, and Interpersonal Comparisons of Utility: A Comment. J Political Econ. 75, 765-766.

\item{[4]} Dillenberger, D., Krishna, R. V. 2014. Expected Utility without Bounds --- A Simple Proof. J. Math. Econ. 52, 143-147.

\item{[5]} Ellsberg, D. 1961. Risk, Ambiguity, and the Savage Axioms. Q. J. Econ. 75, 643-669.

\item{[6]} Foldes, L. 1972. Expected Utility and Continuity. Rev. Econ. Stud. 39, 407-421.

\item{[7]} Gilboa, I., Schmeidler, D. 1989. Maxmin Expected Utility with Non-Unique Prior. J. Math. Econ. 18, 141-153.

\item{[8]} Grandmont, J-M. 1972. Continuity Properties of a von Neumann-Morgenstern Utility. J. Econ. Theory 4, 45-57.

\item{[9]} Harsanyi, J. C. 1955. Cardinal Welfare, Individualistic Ethics, and Interpersonal Comparisons of Utility. J. Political Econ. 63, 309-321.

\item{[10]} Harsanyi, J. C. 1975. Nonlinear Social Welfare Function. Theory Decis. 6, 311-332.

\item{[11]} Harvey, C. M. 1999. Aggregation of Individual Preference Intensities into Social Preference Intensity. Soc. Choice Welf. 16, 65-79.

\item{[12]} Hosoya, Y. 2022. An Axiom for Concavifiable Preferences in View of Alt's Theory. J. Math. Econ. 98, 102583.

\item{[13]} Krantz, D. H., Luce, R. D., Suppes, P., Tversky, A. 1971. Foundations of Measurement vol.1: Additive and Polynomial Representations. Academic Press, New York.

\item{[14]} Kreps, D. 1988. Notes on the Theory of Choice. Westview Press, Boulder.

\item{[15]} Luce, R. D., Raiffa, H. 1957. Games and Decisions: Introduction and Critical Survey. Wiley, New York.

\item{[16]} von Neumann, J. and Morgenstern, O. 1944. Theory of Games and Economic Behavior. Princeton University Press, Princeton.

\item{[17]} Savage, L. J. 1954. The Foundation of Statistics. Dover, New York.

\item{[18]} Schmeidler, D. 1989. Subjective Probability and Expected Utility without Additivity. Econometrica 57, 571-587.

\item{[19]} Sen, A. 1970. Collective Choice and Social Welfare. Elsevier, North-Holland.

\item{[20]} Sen, A. 1976. Welfare Inequality and Rawlsian Axiomatics. Theory Decis. 7, 243-262.

\item{[21]} Shapley, L. S. 1975. Cardinal Utility from Intensity Comparisons. Report R-1683-PR, The Rand Corporation, Santa Monica, CA.

\item{[22]} Weymark, J. A. 1993. Harsanyi's Social Aggregation Theorem and the Weak Pareto Principle. Soc. Choice Welf. 10, 209-221.

\end{description}

\end{document}